\documentclass[twocolumn,showpacs,prl,amsmath,amssymb,floatfix]{revtex4}

\usepackage{epsfig,psfrag}
\usepackage{color}
\usepackage{graphicx}
\usepackage{dcolumn}
\usepackage{bm}
\usepackage{textcomp}

\begin{document}

\title{Microscopic theory of Cooper pair beam splitters based on
carbon nanotubes}

\author{P. Burset$^1$, W. J. Herrera$^2$ and A. Levy Yeyati$^1$}
\affiliation{
$^1$Departamento de F{\'i}sica Te{\'o}rica de la Materia Condensada C-V,
Universidad Aut{\'o}noma de Madrid, E-28049 Madrid, Spain \\
$^2$Departamento de F\'{\i}sica, Universidad Nacional de Colombia,
Bogot\'a, Colombia}
\date{\today}
\begin{abstract}
We analyze microscopically a Cooper pair splitting device in which a central superconducting lead is connected to two weakly coupled normal leads through a carbon nanotube. We determine the splitting efficiency at resonance in terms of geometrical and material parameters, including the effect of spin-orbit scattering. While the efficiency in the linear regime is limited to $50\%$ 
and decay exponentially as a function of the width of the superconducting region we show that it can rise up to $\sim 100\%$ in the non-linear regime for certain regions of the stability diagram. 
\end{abstract}


\maketitle
\begin{figure}[ht!]
\epsfig{file=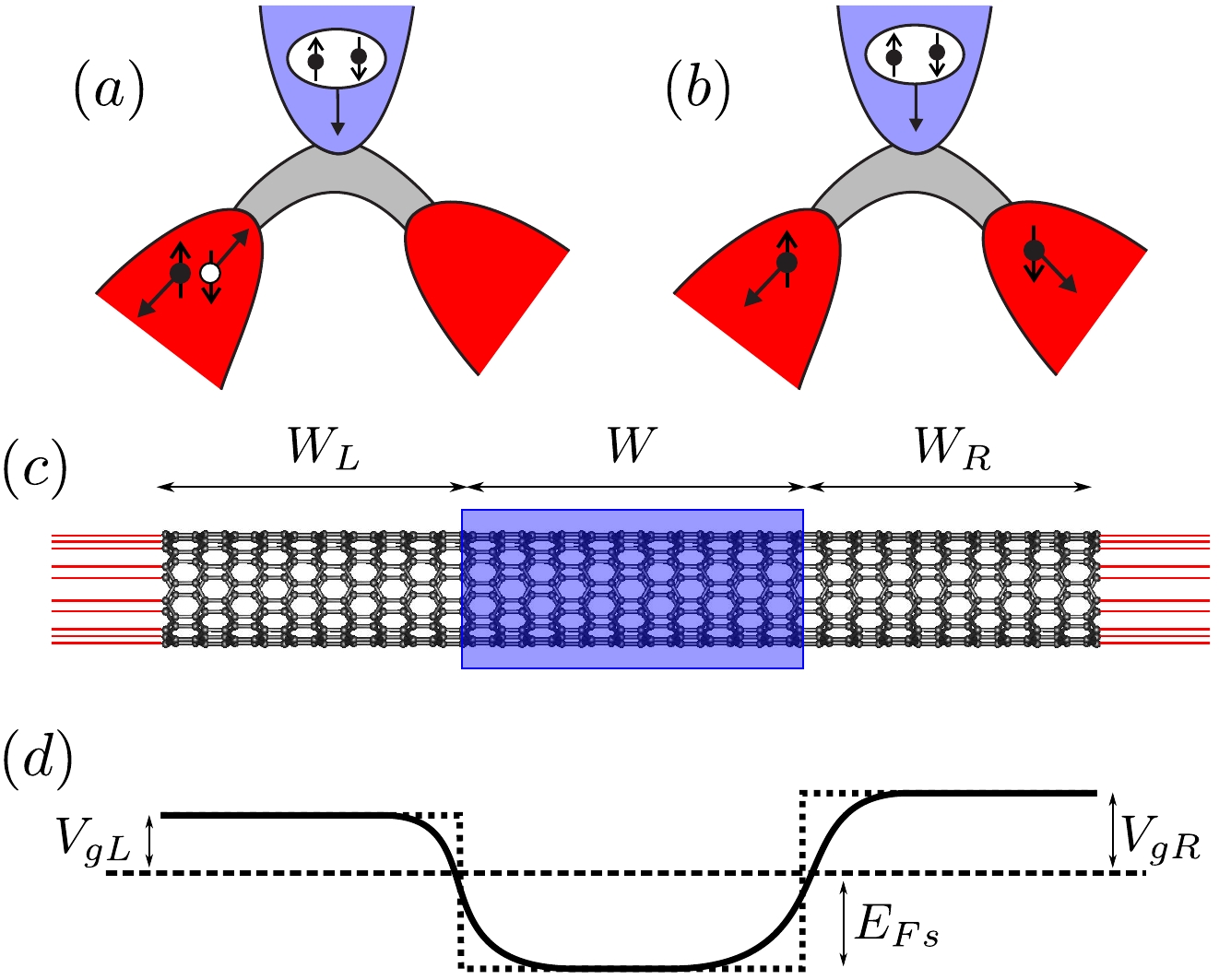,width=8.5cm}
\caption{(Color online) 
Schematic representation of local (a) and non-local (b) Andreev processes in a generic Cooper pair splitter. 
(c) Specific geometry considered in this work: a finite SWCNT coupled to normal leads at its ends and a central superconducting lead. The lower image (d) illustrates the potential profile along the tube.}
\label{figure1}
\end{figure}

{\it Introduction:}
Producing entangled electron pairs in a solid state device
from the splitting of Cooper pairs \cite{general} is a challenging possibility which is starting to generate an intense experimental effort \cite{diffusive,herrmann,hofstatter}. A basic splitting device is a three terminal system with a central superconducting lead (S) in between two normal (N) ones as depicted in the upper panel of Fig. \ref{figure1}. When a Cooper pair is injected from the S lead it can either be transmitted as a whole to one of the N leads by means of a {\it local} Andreev process (Fig. \ref{figure1}a) or split so that each of the electrons in the pair is transmitted to a different lead, which corresponds to a {\it crossed} Andreev process (CAR) (Fig. \ref{figure1}b)
\cite{car}. While initial experimental devices were based on nanolitographically defined diffusive samples \cite{diffusive} more recent experiments are oriented towards tunable double quantum dots systems based either on carbon nanotubes \cite{herrmann} or InAs nanowires \cite{hofstatter}. In spite of the difference in materials the systems realized in both experiments were
conceptually equivalent. They did correspond, however, to different
physical regimes: while in Ref \cite{herrmann}
the hybridization by direct tunneling between the dots was dominant, in Ref. \cite{hofstatter} the direct tunneling appeared to be negligible.
In both works the Cooper pair splitting action was demonstrated indirectly by  
analyzing the changes in the behavior of the conductance when going from the normal to the superconducting state. Both works pointed out an unexpectedly high efficiency for CAR, much higher than what would be predicted by theories which do not take into account the direct inter-dot tunneling \cite{general}.
Still further experimental and theoretical efforts are needed in order to demonstrate the splitting unambiguously and to reach the nearly 100$\%$ efficiency which could be necessary for entanglement detection \cite{bouchiat}.

In the present work we analyze microscopically the case of double quantum dots (DQD) defined on single-walled carbon nanotubes (SWCNTs) and show that the two regimes of Refs. \cite{herrmann,hofstatter} can be reached
in metallic or semiconducting tubes. We consider the situation illustrated in Fig. \ref{figure1}c and \ref{figure1}d where the central electrode modifies the electrostatic potential and induces a pairing amplitude on the portion of the tube underneath without breaking its continuity. In agreement with the experimental observations we show that in this case the splitting efficiency decays rather weakly with the width of the central electrode \cite{hofstatter}. Our results also suggest how to increase the splitting efficiency up to a level close to 100\% by operating the devices in the non-linear regime. We begin with a SWCNT in the normal state without e-e interactions in order to analyze the inter-dot coupling. Subsequently, we switch on superconductivity in the central electrode and study the probability of the CAR processes and the splitting efficiency. Finally, we use a minimal model to analyze the effect of the electron-electron interactions and the non-linear regime, where the efficiency can be highly enhanced.

{\it Basic modeling:}
We focus on zig-zag SWCNTs which allow to consider both metallic and semiconducting cases \cite{mele}. If the coupling to the central lead is sufficiently smooth on the atomic scale we may assume that intervalley scattering is weak and the K-K' degeneracy is preserved. For this case and when the radius is of the order of 1 $nm$ or smaller it is important to consider curvature effects which produce a finite band gap in metallic tubes and enhances the effect of spin-orbit (SO) interactions. We use two complementary approaches for describing the electronic states in the zig-zag SWCNT: a tight-binding (TB) model and a continuous description based on the Bogoliubov-de Genes-Dirac equations. While the latter allows analytical insight, the numerical TB calculations allow to identify effects due to deviations from linear dispersion, disorder or arbitrary spatial variation of the electrostatic potential along the tube.

Within the continuous description the system is characterized by the equations
\begin{equation}
\left( \begin{array}{cc}
H^e_{\tau ,s}-E_{F} & \Delta(x)  \\ 
\Delta(x)  & E_{F}-H^e_{\tau ,s}%
\end{array}%
\right)\! \left( 
\begin{array}{c}
u_{\tau ,s} \\ 
v_{\tau ,s}%
\end{array}%
\right)\! =\! E_{\tau ,s}\left( 
\begin{array}{c}
u_{\tau ,s} \\ 
v_{\tau ,s}%
\end{array}%
\right) 
\label{equation1}
\end{equation}
where 
$H_{\tau ,s}^{e}=-i\hbar v_F \partial_x \cdot \sigma_x + \tau \hbar v_F q_n \sigma_y +\tau
\delta _{0}s-\tau \delta _{1}s\sigma _{y} + V(x)$ is the normal state effective Hamiltonian for the $n$ mode (corresponding to a quantized momenta $q_n$ around the tube), $\Delta(x)$ is the induced pairing amplitude and $V(x)$ the electrostatic potential profile along the tube. In these equations $\sigma_{\mu}$ are Pauli matrices in sublattice space, and $\tau,s = \pm$ correspond to the valley and spin indexes respectively. Finally, the terms in $\delta_0$ and $\delta_1$ take into account the SO interaction as in Refs. \cite{seung,weiss}.
The quantized momenta take the values $q_{n}=\frac{2\pi }{3Na_{0}}\left( n\pm \frac{p}{3}\right) -q_{curv}$, with $p=N \mbox{mod}3=0,\pm 1$ and $N$ being the number of atoms in the cross section. Depending on whether $p=0$ or $p=\pm 1$ the tube is metallic or semiconductor, respectively.
The curvature effect is included in $q_{curv} = E_{curv}/\hbar v_F$, where
$E_{curv} \simeq \pi^2 |V_{pp\pi}|/4N^2$ with $|V_{pp\pi}| \simeq 2.7 eV$. In the normal homogeneous case the corresponding energy levels for longitudinal wavevector $k$ are thus given by $E^n_{\tau,s}(k) = \hbar v_F \sqrt{(q_n + \tau s \delta_1)^2 + k^2} + \delta_0 \tau s$.
The transport properties can be expressed in terms of Green functions which satisfy $(E - {\cal H}_{\tau,s}(x)) G_{\tau,s}(x,x')
= \delta(x-x')$, where ${\cal H}_{\tau,s}(x)$ denotes the full Hamiltonian on the left hand side of
Eq. (\ref{equation1}). We obtain these quantities by solving first for the uniform finite regions and then 
matching the result using the method of Ref. \cite{herrera}. 

\begin{figure}[ht!]
\epsfig{file=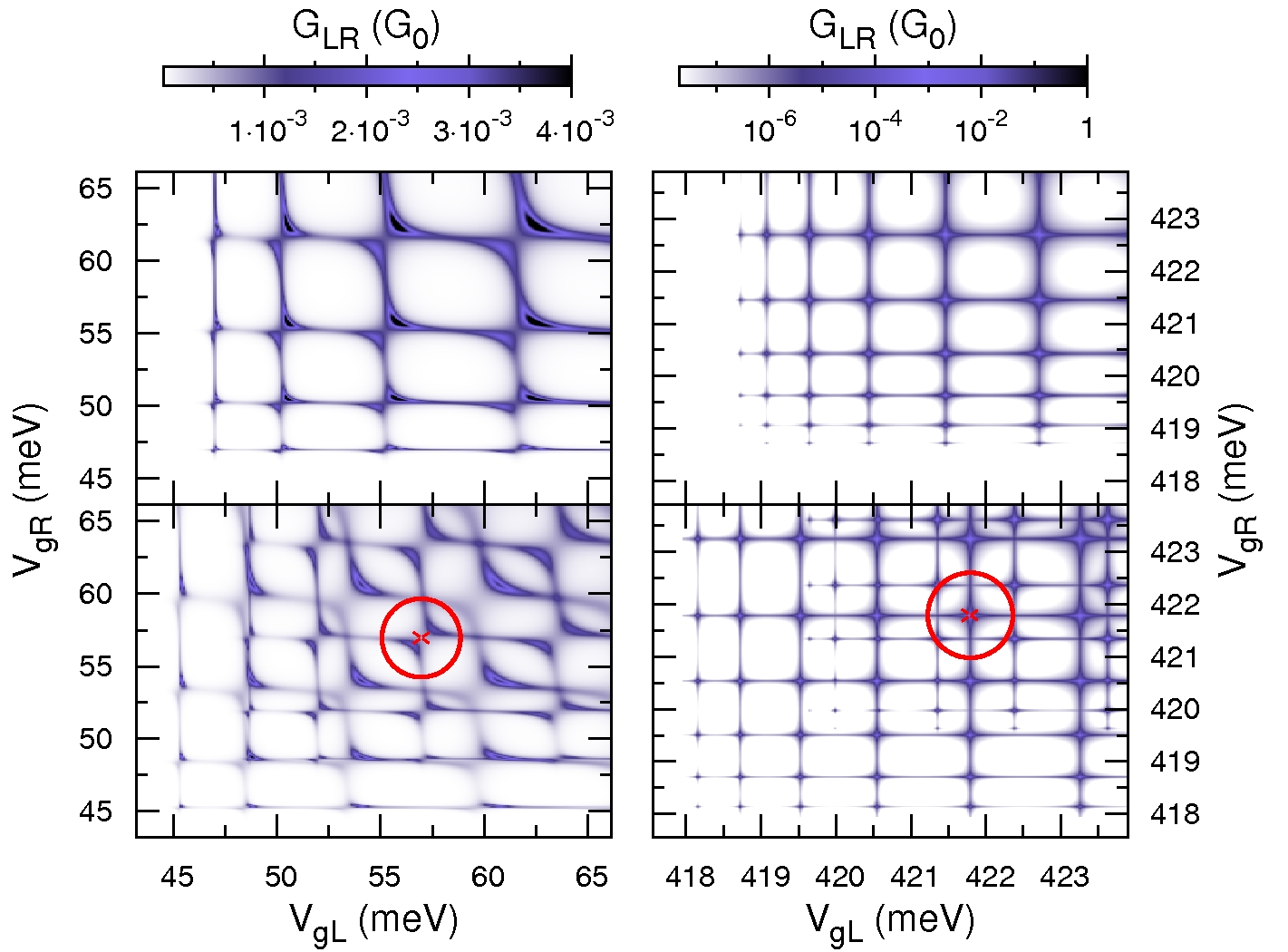,width=8.5cm}
\caption{(Color online) 
Left panel: Conductance map for a metallic tube ($N=12$) in the normal state
for the p-n-p region with (lower) and without (upper) SO interactions.
Right panel: Same for the semiconducting tube ($N=11$) but in a logarithmic scale.
The geometrical parameters are $W=W_{L,R}=170nm$.}
\label{figure2}
\end{figure}

{\it Normal state:} 
We start by analyzing the linear conductance along the tube, $G_{LR}$, when the central lead is in the normal state. In Fig. \ref{figure2} we show a map of $G_{LR}$ in the $V_{gL}-V_{gR}$ plane, obtained using the TB model in the usual nearest neighbors approximation with a hopping parameter $t\equiv V_{pp\pi}$. As a first approximation the potential profile along the tube is assumed to change discontinuously as represented by the dashed lines in Fig. \ref{figure1}d. The lateral leads are modeled by ideal 1D channels weakly coupled to the ends of the tube, as represented schematically in Fig. \ref{figure1}c. We fix the tunneling rates to these leads to a value $\Gamma_{L,R} \sim 0.01 t$ which is consistent with the conductance values observed in Ref. \cite{Kuemmeth} for the lowest energy states of a SWCNT quantum dot. To model the effect of the central lead we rely on the results of ab-initio calculations for the case of Al electrodes \cite{giovannetti,barrasa-lopez}. According to Ref. \cite{giovannetti} these produce a n-doping effect, leading to a shift of the tube bands $E_{Fs} \sim -0.5 eV$ for an ideal interface. On the other hand in the normal state it would induce a broadening of the tube levels of the order of a few $meV$ \cite{barrasa-lopez} which suggests a typical value $\Gamma_S \sim 1 meV$ for the corresponding tunneling rate. As in the experiments of Ref. \cite{herrmann} the length of the central region is set initially to $\sim 200 nm$. We consider tubes with $N=11,12$ which corresponds to radii $R \sim 0.43,0.47 nm$ for the semiconducting and metallic cases respectively. In the metallic case curvature effects lead to the opening of a narrow gap, which can be estimated as $E_g\simeq E_{curv} \simeq 45 meV$. The curvature gap is apparent in the upper left panel of Fig. \ref{figure2}. On the other hand, in the semiconducting case the gap is $E_g \simeq 412 meV$ (top right panel of Fig. \ref{figure2}). It should be noticed that for these diameters and for gate potentials of the order of $0.5 eV$ only the lowest energy mode, corresponding to $n=0$, gives a significant contribution to the transport properties of the tubes that we discuss below.

For positive $V_{gL},V_{gR}$, i.e. in the p-n-p  regime the conductance displays a DQD behavior as shown in Fig. \ref{figure2}.
The metallic case (left panels) exhibits an anticrossing pattern similar to the one found in the experiments of Ref. \cite{herrmann}.
As the gate potentials $V_{gL},V_{gR}$ become increasingly positive the conductance map exhibits resonances along lines $V_{gL} + V_{gR} 
\sim const$ indicating the delocalization of the electronic states due to Klein tunneling. The confinement of the dot states is much more pronounced in the semiconducting case where the Klein tunneling is less significant. We have used a logarithmic scale in this case in order to enhance the visibility of the conductance peaks.

When SO scattering is introduced (lower panels in Fig. \ref{figure2})
there is a general splitting of the conductance peaks of the order of $\sim 2 meV$ due to the breaking of the spin-valley degeneracy. Close to the gap edges this splitting is of the same order as the mean level separation.

{\it Superconducting state:}
When superconductivity in the central lead is switched-on pairing correlations within the tube are induced by proximity effect. The size of the induced gap $\Delta_{i}$ is set by $\Gamma_S$ (i.e. of the order of $1meV$).
We shall assume that temperature is zero and that the energy $E$ of the injected 
electrons from the normal leads is smaller than $\Delta_{i}$. 
Then $R_{AL}(E)$ and $R_{AR}(E)$ denote the local Andreev reflection probabilities at the $L,R$ leads while $T_{CAR}(E)$ correspond to the CAR processes. When operated as a beam splitter a finite voltage difference $V$ is applied between the S and the N leads and the non-linear conductance is given by 
$G_{L(R)}(V) = G_0 (T_{CAR}(V)+T_{CAR}(-V)+2 R_{AL(R)}(V))$, with $G_0 = 2e^2/h$ (notice that
at finite energy in general $T_{CAR}(E) \ne T_{CAR}(-E)$ due to the breaking of the electron-hole symmetry).  
Thus we can define the splitting efficiency as $\eta=G_0 (T_{CAR}(V)+T_{CAR}(-V))/(G_L(V) + G_R(V))$.

\begin{figure}[ht!]
\epsfig{file=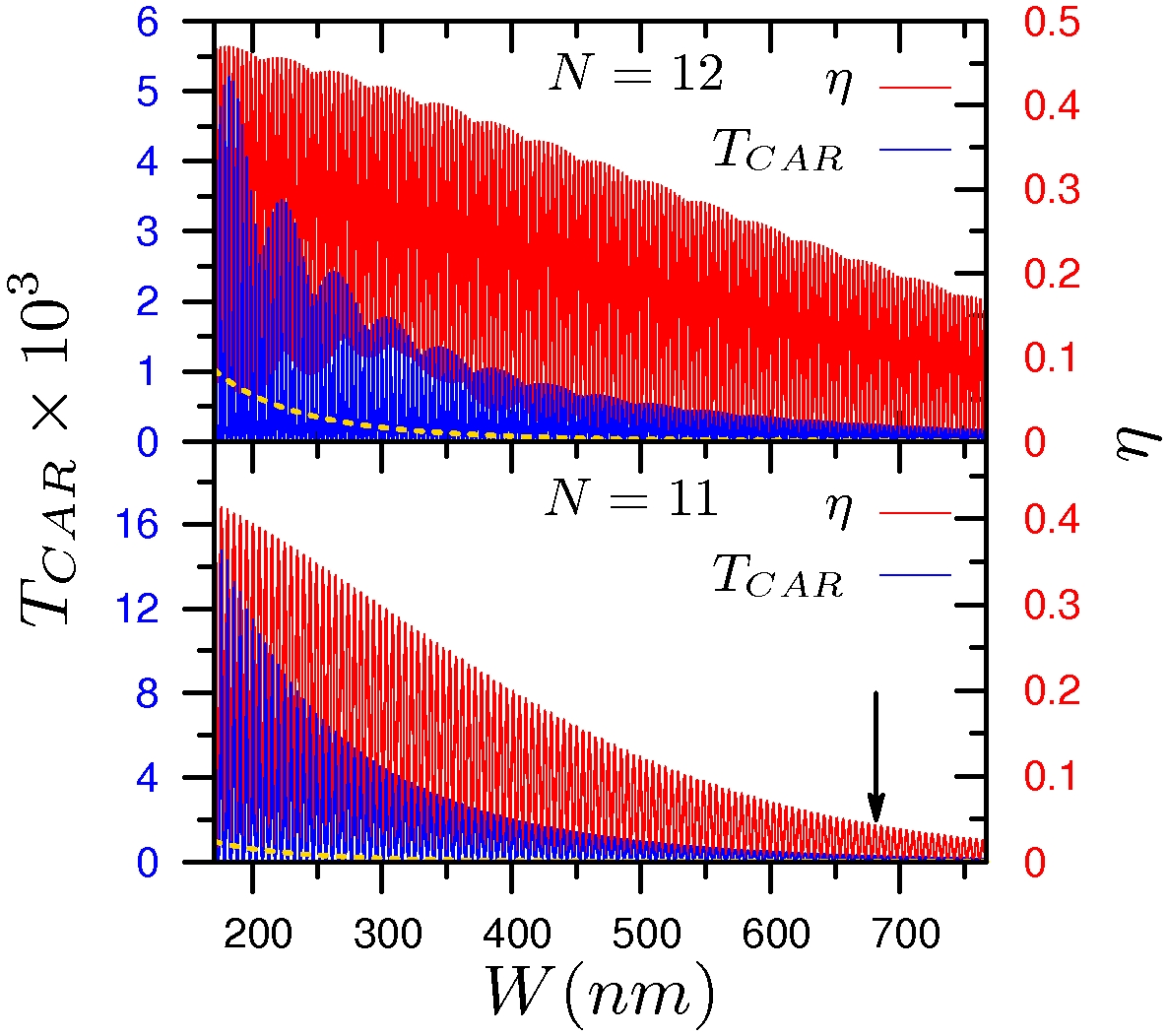,width=8.5cm}
\caption{(Color online) 
Evolution of the CAR probability (blue) and the splitting efficiency $\eta$ (red) in the linear regime
as a function of the length $W$ of the central electrode for a metallic SWCNT with $N=12$ (upper panel) and a semiconducting one with $N=11$ (lower panel). The gate potentials $V_{gL,gR}$ are fixed at the points indicated by the 
circles in Fig. \ref{figure2}. The dashed lines represent the decay of the CAR probability for a 3D bulk superconductor multiplied by a factor $1000$.}
\label{figure3}
\end{figure}

Within both the Dirac and TB models it is found that the CAR coefficients decay exponentially on the scale $\xi(q) = \hbar v_F/\Delta_{i}\sqrt{1 - (\hbar v_f q/E_{Fs})^2}$, where $q = q_0 \pm \delta_1/\hbar v_F$ exhibiting oscillations on 
the scale $\lambda_F = \hbar v_F/|E_{Fs}|$, as illustrated for the linear regime, $V=0$, in Fig. \ref{figure3}. In these plots we have fixed the gate voltages at the values indicated by the circles in Fig. \ref{figure2}.  
The CAR probability decays more slowly in the metallic case ($N=12$) due to the longer effective coherence length $\xi(q)$ of this case. In both cases, however, the decay is remarkably slower than in the case of a 3D bulk BCS superconductor where the prediction is $\sim \exp{(-2W/\xi)}/(k_F W)^2$ \cite{decay-3D}, indicated by the dashed lines in Fig. \ref{figure3}. This is a consequence of
the single channel character of the connection between the dots in the present system and explain the rather large efficiency values estimated in recent experiments \cite{hofstatter}. As can be observed the efficiency $\eta$ decreases from 0.4 at $W \sim 200 nm$ to $<0.05$ at $W \sim 700 nm$ in the semiconducting case, while it varies between 0.5 and 0.2 for the metallic tube within the same $W$ range. For sufficiently large $W$ the overall evolution of $\eta$ is well described by the expression $\eta \sim 1/(1 + \exp{2W/\xi(q)})$. This qualitative behavior is also found for a smoother potential profile \cite{SI}.

{\it Coulomb interactions and non-linear regime:}
For analyzing the effect of electron-electron interactions we first map the 
system into a minimal model in which we keep just one two-fold degenerate electron level $E_{L,R}$ in each dot \cite{comment}. In the combined dot-Nambu space the model properties can be expressed in terms of bispinor fields $\Psi_{\mu}\!\! =\!\! (d_{\mu\uparrow}, d^{\dagger}_{\mu\downarrow})$ where $\mu\!=\!L,R$ and $d^{\dagger}_{\mu\sigma}$ creates dot electrons. In the absence of interactions this reduced model is characterized by a retarded Green function matrix of the form $\hat{G}^{(0)}\!\! =\!\! \left[ E\! -\! \hat{h}_0 \! + \! i\hat{\Gamma}\! -\! \hat{\Sigma}(E) \right]^{-1}$, where $(\hat{h}_0)_{\mu\nu,\alpha\beta}\! =\! E_{\mu} \delta_{\mu\nu} \delta_{\alpha\beta} (-1)^{\alpha+1}$, with $\mu,\nu \equiv L,R$, $\alpha,\beta \equiv 1,2$ are the Nambu indexes, $(\hat{\Gamma})_{\mu\nu,\alpha\beta}\! =\!
\tilde{\Gamma}_{\mu} \delta_{\mu\nu}\delta_{\alpha\beta}$, with $\tilde{\Gamma}_{\mu}\!=\! \Gamma_{\mu}a_0/W_\mu$, correspond to the effective tunneling rates to the normal leads and $\hat{\Sigma}$ is a matrix self-energy describing the coupling with the central superconducting region. The whole analytical expression for $\hat{\Sigma}$ is given in the supplementary information.

Interactions are introduced by assuming a constant charging energy $U_{L,R} \gg \Delta_{i}$ acting on each dot level. We take them into account within the equation of motion (EOM) technique with a Hartree-Fock decoupling at the level of the two-body Green functions \cite{SI}. This approximation is valid when Kondo and exchange correlations between the dots can be neglected. Further simplification is achieved in the limit $U_{\mu} \rightarrow \infty$ where we find $\hat{G}\!\! =\!\! \left[\hat{g}^{-1}\! +\! i\hat{\Gamma}\! -\! \hat{\Sigma}\right]^{-1}$,
with $g\!\! =\!\! (E\! -\! \hat{h}_0)^{-1} [1\! -\! \hat{A}_{\infty}]$ and $(\hat{A}_{\infty})_{\mu\nu,\alpha\beta}\! =\! n_{\mu} \delta_{\mu\nu} 
\delta_{\alpha\beta}$. The evaluation of the mean values
$n_{\mu}\! = <\!\!d^{\dagger}_{\mu\sigma}d_{\mu\sigma}\!\!>$ must be
performed self-consistently. The relevant transport coefficients are finally computed as $T_{CAR}(E)\! =\! 4 \tilde{\Gamma}_L\tilde{\Gamma}_R |G_{LR,12}(E)|^2$ and $R_{AL(R)}\! =\! 4 \tilde{\Gamma}_{L(R)}^2 |G_{LL(RR),12}|^2$.

\begin{figure}[ht!]
\epsfig{file=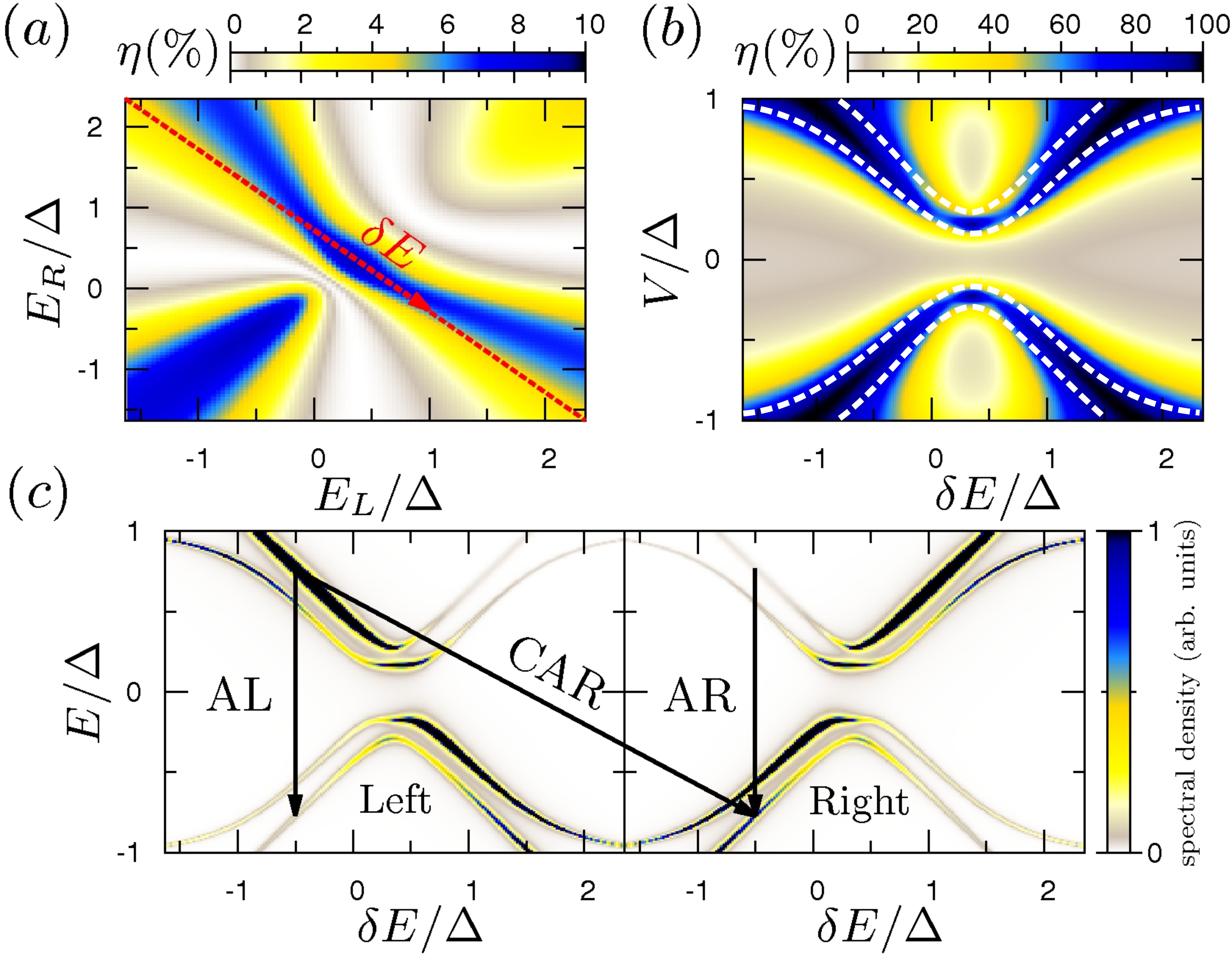,width=8.5cm}
\caption{(Color online) Color map of the splitting efficiency within the minimal model with parameters corresponding to a semiconducting tube with $W\sim 700 nm$ (indicated by the arrow in Fig. \ref{figure3}) in the linear (a) and non-linear (b) regimes. In the latter case the dot levels are varied along the line $E_L \sim -E_R$ indicated by the dashed red line of (a). The white dashed lines indicate the maxima in the spectral density which is shown in (c) for the two dots along this line. Local and non-local Andreev processes at finite $V$ are indicated by the black arrows.
}
\label{figure4}
\end{figure}

The main effect of interactions within this approximation is to shift the resonances and to reduce their width, roughly as $(1-n_{\mu}) \tilde{\Gamma}_{\mu}$. Then, the CAR and the local Andreev probabilities are reduced by a factor $(1 - n_1)(1 - n_2)$ and $(1 - n_{\mu})^2$ respectively, which therefore does not modify significantly the efficiency at resonance. The color map in Fig. \ref{figure4}a shows the efficiency in the linear regime corresponding to the semiconducting case with $W \sim 700 nm$ (arrow in Fig. \ref{figure3}) and for the region of gate voltages indicated by the circle in the right panel of Fig. \ref{figure2}. As can be observed, $\eta$ exhibits maximum values at the crossing point between the resonances of the order of $0.1$ which are slightly higher than the values found in the non-interacting case. The efficiency reaches a maximum of the same magnitude along the line $E_L \sim -E_R$ (red dashed line). What is much more remarkable is that the efficiency along this line can rise up to 100\% in the non-linear regime $V \neq 0$. This is illustrated in Fig. \ref{figure4}b. The high efficiency regions lie within the dot resonances (indicated by the dashed white lines) which are shifted from zero energy due to the presence of an induced minigap by the proximity with the superconducting lead.

The origin of these high efficiency regions can be understood qualitatively from the spectral density on each dot, which is shown in Fig. 
\ref{figure4}c. At any given point along the line $E_L \sim -E_R$ electron and hole states are split due to hybridization between the dots. 
Furthermore, the electron-hole symmetry in the local spectral density is lost along this line. 
Crossed Andreev processes like the one sketched as the CAR arrow in Fig. \ref{figure4}c combines electron and hole states on each dot 
with high spectral density. 
These inter-dot transitions are then more favorable than the intra-dot electron-hole conversions (arrows AR and AL), in which either the electron or 
the hole state has low spectral density. As a consequence, local Andreev processes become suppressed while non-local CAR processes are enhanced, 
thus explaining the efficiency increase.

{\it Conclusions:} We have analyzed the splitting efficiency of SWCNT double quantum dot devices in terms of material and geometrical parameters. The single channel character of the connection between the dots in this configuration explains the weak decay of CAR with distance which is consistent with the experimental observations. Furthermore we have shown how the splitting efficiency can rise up to $\sim 100\%$ by working in the non-linear regime.
We expect that our analysis can guide future experiments for the production of entangled electron pairs using these devices .

The authors would like to thank T. Kontos, A. Cottet and R. Egger for fruitful discussions. This work was supported by MICINN-Spain via grant FIS2008-04209 (PB and ALY) and DIB from Universidad Nacional de Colombia, project 12170 (WJH).


\begin{thebibliography}{99}
\bibitem{comment} The effect of SO interactions is to break the four-fold degeneracy and is thus implicitly taken into account in the minimal model.
\end{thebibliography}
\end{document}